\documentclass[12pt, draftclsn, onecolumn]{IEEETran}
\renewcommand{\baselinestretch}{2}
\newcommand{\CLASSINPUTinnersidemargin}{1in}
\newcommand{\CLASSINPUToutersidemargin}{1in}

\usepackage{amsmath}
\usepackage{amsfonts}
\usepackage{epsfig}
\usepackage{amssymb}
\usepackage{algorithmic}
\usepackage[ruled]{algorithm}

\begin{document}

\title{\huge{Sensing Matrix Setting Schemes for Cognitive Networks and Their Performance Analysis}\thanks{ }}

\author{\authorblockN{Hossein Shokri-Ghadikolaei, Masoumeh Nasiri-Kenari \\{\small Wireless Research Lab., Elec. Eng. Dept., Sharif University of Technology}}}

\maketitle
\begin{abstract}
Powerful spectrum decision schemes enable cognitive radios (CRs) to
find transmission opportunities in spectral resources allocated
exclusively to the primary users. One of the key effecting factor on the CR network throughput is the spectrum sensing sequence used by each secondary user. In this paper, secondary users' throughput maximization through finding an appropriate sensing matrix (SM) is investigated. To this end, first the average throughput of the CR network is evaluated for a given SM. Then, an optimization problem based on the maximization of the network throughput is formulated in order to find the optimal SM. As the optimum solution is very complicated, to avoid its major challenges, three novel sub optimum solutions for finding an appropriate SM are proposed for various cases including perfect and non-perfect sensing. Despite of having less computational complexities as well as lower consumed energies, the proposed solutions perform quite well compared to the optimum solution (the optimum SM). The
structure and performance of the proposed SM setting schemes are
discussed in detail and a set of illustrative simulation results
is presented to validate their efficiencies.
\end{abstract}


\section{Introduction}

\PARstart{O}{pportunistic} spectrum sharing has been developed
through the new promising concept of Cognitive Radio (CR), in
order to meet ever-growing spectrum demands for new wireless
services. Conceptually, CR is an adaptive communication system
which offers the promise of intelligent radios that can learn from
and adapt to their environment \cite{Clancy07}. The major issue in
designing a cognitive radio network is to protect
incumbent/primary users from potential interference problems while
providing acceptable quality-of-service (QoS) levels for secondary
users (i.e., unlicensed users). To this end, sensing capability is
exploited in CRs which enable them to find some transmission
opportunities called white spaces, i.e., temporarily-available spectrums which are not used by primary users (PUs). Limited number of possible observations and dynamic
nature of observed signals lead to imperfect sensing
which is usually described by false alarm and miss detection probabilities. The false alarm occurs when
the PU is idle, but the Secondary User (SU) senses the channel
as busy. While the miss detection is occurred when the SU senses
an occupied channel as free.

Average throughput of the SUs is one of the most important
performance metrics which depends on the candidate primary
channels for sensing and transmission, and it must be considered in
designing appropriate sensing schemes. Generally, there exists
more than one channel to be sensed by a CR. As a result,
sensing schemes are commonly divided into two categories,
i.e., wideband sensing and narrowband sensing. Sensing is wideband
when multiple channels are sensed simultaneously. These multiple
sensed channels can cover either the whole or a portion of the
primary channels \cite{pedram2011}. On the other hand, when only
one channel is sensed at a time, the sensing is
narrowband. Ease of implementation, lower power consumption, and
less computational complexity lead to great interest in
narrowband sensing. When the narrowband sensing is used, an
immediate question arises: which channel should be sensed first?
In other words, to achieve the best possible performance, the
channels have to be sensed in an appropriate order determined by sensing sequence (SS).


In \cite{zhao2007}, the problem of joint optimization of sensing and transmission is
addressed. Specifically, Zhao et
al. in \cite{zhao2007} proposed a decentralized slotted CR MAC
protocol to grasp the optimal policies for spectrum sensing and
access framework through a partially observable Markov decision process.
Minimizing the overall system time of a SU, which contains the average waiting time and the extended data delivery time, through load balancing in probability-based and sensing-based spectrum decision schemes is investigated in \cite{wang11}.
In \cite{liu08}, \cite{win08}, and \cite{hamdaoui09}, the procedures to determine the optimal set of candidate channels for sensing are first discussed and then the maximization of the spectrum accessibility through optimal number of
candidate channels are investigated. In \cite{lai08}, \cite{sabharwal07}, and
\cite{jia08}, the sequential channel sensing problems are formulated
based on maximizing the throughput
of the SUs. While in these works, the optimum sensing times have been studied, the effects of the sensing errors have not been addressed.
Setting a SS by prioritizing the various channels can play a major role in finding a transmission
opportunity or equivalently expected SU's throughput.
Channel prioritization has been considered in \cite{chang2007} in which an optimal channel sensing framework for a single-user case including the sensing order and
the stopping rule has been proposed. In \cite{chang2007}, it has been also
assumed that the SUs are allowed to recall and guess. Recall means
the ability to go back and access a previously sensed channel and
guess means accessing a channel that has not been sensed yet. In \cite{chow1971} and \cite{zheng09}, a stopping rule has been developed to determine when to stop the sequential sensing
procedure and when to start secondary transmission.
In \cite{kim08}, the optimal SS has been derived for channels with
homogeneous capacities, and it was shown that the problem of
finding the optimal sensing sequence for these channels
is NP-hard. The authors in \cite{Kim2008}
have suggested a SS which sorts channels in descending order
according to their idle probabilities. In \cite{jiang2009}, finding the
optimal SS sequence has been investigated for a single-user
case with the aim to maximize the SU's throughput. The problem of finding optimal SS for two CR
users has been addressed in \cite{fan2009}, in which an exhaustive search has been applied in order to find the
best sensing sequences for the users at the expense of a huge computational
complexity. To reduce the complexity associated with the optimum solution, the authors of \cite{fan2009} have proposed two low-complexity
suboptimal algorithms with the achieved throughput close to
the maximum possible value.

In this paper, the problem of selecting proper spectrum sensing
sequences for a cognitive radio network (CRN) with multiple users is addressed.
Our objective is to maximize the average network throughput. First, we assume a perfect
channel sensing (i.e., error-free sensing) and formulate an
optimization problem on spectrum sensing sequences of the SUs based on
maximizing the average throughput of the network. We discuss the
conventional solution as well as its computational
complexity. Due to massive computational burden of the conventional
optimization algorithm, a novel algorithm, which finds
near-optimal solution, is proposed. The proposed algorithm, called
sensing matrix setting (SMS), provides short-term and mid-term
fairness among the SUs and offers near-optimal solution with
tolerable computational complexity as well as relatively low consumed
energy. Then, we
consider the impact of sensing error on the SMS algorithm, and
propose modified version of SMS algorithm, called MSMS algorithm.
In addition, for the multiple access among the SUs, we apply the
conventional p-persistent MAC within the MSMS
algorithm, and call the extended algorithm as PMSMS algorithm. Structure,
performance, and related spectrum allocation processes for
the proposed algorithms are discussed in detail.

The rest of this paper is organized as follows. In Section \ref{section2}, we
describe the CR network considered and the related assumptions.
In Section \ref{section2}, the throughput of the CRN for a given sensing
matrix is formulated, and the conventional approach to find the optimal
SM as well as its computational complexity are discussed. In
Section \ref{section4}, the structure, computational complexity, and consumed
energy of the novel suboptimal SMS algorithm are described in detail. In
Section \ref{section5}, the modified version of the SMS algorithm is introduced.
The PMSMS algorithm is described in Section \ref{section6}. Numerical results
are then presented in Section \ref{section7}, which validate our analysis and
verify the advantages of the proposed algorithms. Finally, the paper
is concluded in Section \ref{section8}.

\section{System Model}\label{section2}

We assume a time slotted CRN with $N_s$ secondary users which
attempt to opportunistically transmit in the channels dedicated to the
$N_p$ PUs. As in \cite{zheng09}, \cite{jiang2009}, \cite{fan2009},
and \cite{liang2008}, the SUs are time synchronous in time-slots
with other SUs and with the PUs. When a PU has no data for
transmission, it does not use its time-slots; and hereby provides a
transmission opportunity for the SUs. That is, at the beginning of each time-slot, a channel
can be established as occupied or vacant. In order to find the transmission opportunities
appropriately and to protect the PUs from harmful interference,
the sensing process is performed at the beginning of each
time-slot. We assume that the SUs are equipped with simple
transceivers, so they are able to sense only one channel at a time.
The SUs always have packets to transmit, and as a consequence they will start
transmission when an opportunity is found. Each SU senses the channels according to its SS
sequentially, i.e., the SU senses the first channel from the top of
its SS for a predetermined time duration (channel sensing time),
and then senses the second channel if and only if the first
channel is found busy. This procedure will continue until a
transmission opportunity is found. Moreover, as \cite{jiang2009}, we assume that the
SUs are not able to "recall" which means that they cannot re-sense and transmit
on a previously sensed busy channel.

The SU might stop its transmission in a channel and try to choose a new one due to the presence of the PU or the availability of a better channel with a more appropriate transmission condition.
In order to switch to a new channel, which is called spectrum Hand-Over (HO), a secondary
device needs a specific and constant time duration $\tau_{ho}$ to
prepare its sensing circuitry for the next spectrum sensing.

For the SU, each slot contains two phases: 1) sensing phase, and
2) transmission phase. The sensing phase contains several
mini-slots of duration $\tau$ (sensing time of each channel).
Sensing is carried out by the SUs in the mini-slots, and once the
transmission opportunity is found, the transmission phase will be
started. This kind of access, i.e., listen-before-talk (LBT), is a
common method in many wireless communication systems, for example see the
quiet period in \emph{IEEE 802.22} standard \cite{stevenson}. The
sensing procedure is performed in an order based on the SS
provided by the secondary network coordinator.
The SUs do not have the adaptive modulation and coding (AMC) capability; so they
transmit with a constant rate,
$R$, during the transmission phase. We define a sensing matrix (SM) as
a matrix with the dimensions of ${N_s} \times {N_p}$, in which the
$i$-th row contains the SS dedicated to the $i$-th SU. Given the
primary-free probabilities, i.e., ${P_{0,j}}{\rm{~}},{\rm{~}}1 \le
j \le {N_p}$, and predetermined false alarm and detection
probabilities, our objective is to find the optimal (or
near-optimal) SS of each SU, i.e., the optimal SM, in order to
maximize the CRN throughput.

\begin{figure}[h]
  \centering
  \includegraphics[width= 5 in]{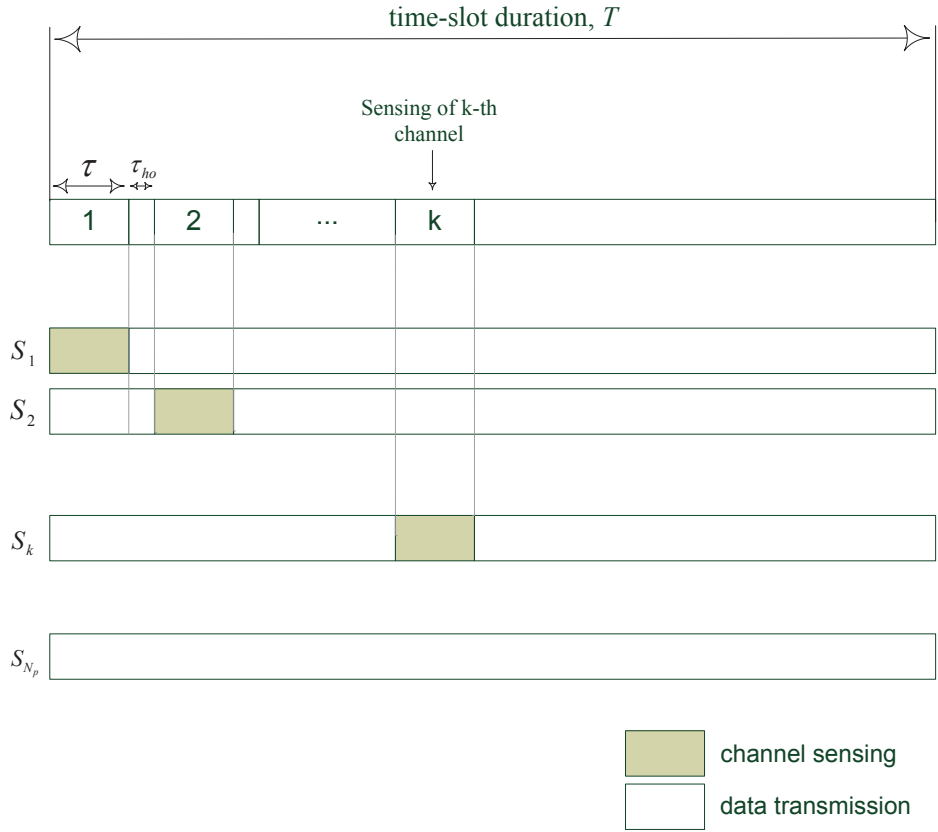}\\
  \caption{General timing structure of our system model.}
  \label{fig1}
\end{figure}

Fig.~\ref{fig1} demonstrates the slotted timing structure of a SU
and its sensing process. For the example considered in Fig.~\ref{fig1}, after
sensing $ \left( k - 1 \right)$ occupied channels, the SU senses
the $k$-th channel free and then transmits data on that channel
until the end of the slot. The wasted time length, i.e., the time
allocated for sensing and HO in this process is equal to $ \tau +
\left( k - 1 \right) \left( \tau + \tau_{ho} \right) $. Thus, the time
left in the slot for transmission is $ T - \tau - \left( k
- 1 \right) \left( \tau + \tau_{ho} \right) $, where $T$ is the
time slot duration. Generally speaking, each slot is composed of a
sensing phase with the maximum length of $ \tau + \left( N_p - 1
\right) \left( \tau + \tau_{ho} \right) $ and a transmission phase
with the minimum length of $ T - \tau - \left( N_p - 1 \right)
\left( \tau + \tau_{ho} \right) $. Let us define the slot effectiveness, denoted by $e_s$, as the ratio of the transmission phase length to the slot length. Hence, if a secondary user starts transmitting on
the $k$-th channel of its SS, the slot effectiveness is:
\begin{equation}\label{eq1}
e_s = \frac{transmission~time}{time~slot~duration} = 1-\frac{\tau + \left(
k - 1 \right) \left( \tau + \tau_{ho} \right)}{T}
\end{equation}
and
\begin{equation}\label{eq2}
 B_k = R \times e_s = R \left( 1-\frac{\tau + \left( k - 1 \right)
\left( \tau + \tau_{ho} \right)}{T} \right)
\end{equation}
where $B_k$ is the average throughput of the SU if the $k$-th
channel of the SS is sensed to be free and chosen for the transmission. From (\ref{eq1}) by the increase of $k$, the channel effectiveness is reduced.

\section{Optimal Sensing Matrix}\label{section3}

In this section, we evaluate the CRN throughput for error-free
sensing case and discuss about the optimal sensing sequences of the SUs (or equivalently SM) for
the network throughput maximization. The optimal sensing sequence for
the CRN containing just one SU is derived in \cite{Kim2008}. For that case, all required was to sort the spectrums based on their primary-free probabilities (i.e., the absence probability of\
the PU). But in the CRN with multi users, the impact of collisions among the SUs' transmissions has to be taken into account. Assume that $S$ denotes the sensing matrix which contains $N_s$ rows and $N_p$ columns with the element $s_{i,j}$ indicating that the $i$-th SU senses the $s_{i,j}$-th spectrum in its $j$-th
mini-slots.

For network throughput evaluation, we note that for each spectrum $s_{i,j}$ two possible cases might occur. First, the
$s_{i,j}$-th spectrum has been sensed by some SUs in their previous mini-slots (the mini-slots before $j$-th mini-slot). In this case, regardless of the presence or the absence of the PU, the spectrum is occupied (as we assumed error-free sensing in this section). That is, the SU that senses this channel
at the first time will transmit on the channel, as a result of perfect
sensing, if the spectrum is idle. Second, the spectrum is sensed at the
$j$-th mini-slot by $i$-th SU at the first time. So, the
occupation probability of this channel only depends on the PU
activity. If the $s_{i,j}$ is sensed free, the $i$-th SU
starts its transmission in this spectrum for the rest of the time
slot with a constant rate $R$. From the above discussion, repetition of  a spectrum in the sensing matrix when assuming error-free sensing does not offer any benefit to the CRN throughput.

Fig.~\ref{fig2} shows the structure of the SM. In this Fig., $Z_{j-1}$ demonstrates the spectrums allocated prior to the $j$-th mini-slot. The array $Y$ contains the spectrums dedicated at the $j$-th mini-slot to
the SUs prior to the user $i$. 
Let $A_{{s_{i,j}}}^{{Z_{j-1}}}$ indicate the presence or absence of the spectrum $s_{i,j}$ in $Z_{j-1}$,
\begin{equation}\label{eq3}
A_{{s_{i,j}}}^{{Z_{j-1}}} = \left\{ {\begin{array}{*{20}{c}}
   {0{\text{~:~}}if{\text{~}}{s_{i,j}} \in {Z_{j-1}}} \hfill  \\
   {1{\text{~:~}}if{\text{~}}{s_{i,j}} \notin {Z_{j-1}}} \hfill  \\
\end{array} } \right.
\end{equation}

By the above definition, when not taking into account the impact of collision
caused by multi SU transmissions at the same spectrum due to simultaneously finding the spectrum free at the $j$-th mini-slot, the spectrum $s_{i,j}$ can be efficiently used by the
$i$-th SU with the probability of
${P_{0,{s_{i,j}}}}A_{{s_{i,j}}}^{{Z_{j - 1}}}$, where $P_{0,{s_{i,j}}}$ is the absence probability of the $s_{i,j}$-th PU. To consider the
impact of the collision on the network throughput as discussed above, we define the
operator $ \oplus $ as follows.
\begin{equation}\label{eq4}
\oplus :\left\{ {\begin{array}{*{20}{c}}
   {A \oplus B = B \oplus A} \hfill  \\
   {A \oplus B \oplus C = A \oplus C \oplus B = B \oplus C \oplus A} \hfill  \\
   {\forall m \leqslant {N_p}:{P_{0,{s_1}}}{A_{{s_1}} ^{{Z_{j-1}}}} \oplus {P_{0,{s_2}}}{A_{{s_2}} ^{{Z_{j-1}}}} \oplus  \cdots  \oplus {P_{0,{s_m}}}{A_{{s_m}} ^{{Z_{j-1}}}} = \mathchar'26\mkern-10mu\lambda } \hfill  \\
\end{array} } \right.
\end{equation}
where
\begin{equation}\label{neql4}
\mathchar'26\mkern-10mu\lambda  = \left\{ {\begin{array}{*{20}{c}}
   0 & {{\text{iff~}}{s_1} = {s_2} =  \cdots  = {s_m}}  \\
   {{P_{0,{s_1}}}{A_{{s_1}} ^{{Z_{j-1}}}} + {P_{0,{s_2}}}{A_{{s_2}} ^{{Z_{j-1}}}} +  \cdots  + {P_{0,{s_m}}}{A_{{s_m}}} ^{{Z_{j-1}}}} & {{\text{iff~}}{s_1} \ne {s_2} \ne  \cdots  \ne {s_m}}  \\
\end{array} } \right.
\end{equation}

The operator $ \oplus $ is used to model the possible collision due to multi SUs finding the same spectrum free at the $j$-th mini-slot. As stated before, each channel in each time-slot has
a contribution in the whole throughput if and only if it is sensed
only by one SU (i.e., assigned to one SU sensing sequence) because
of error-free sensing assumption.

By the above definition of the operator $ \oplus $, the average throughput of the
CRN is easily computed as follows,
\begin{equation}\label{eq5}
\begin{gathered}
  Q = \left( {{P_{0,{s_{1,1}}}}A_{{s_{1,1}}}^{{Z_0}} \oplus {P_{0,{s_{2,1}}}}A_{{s_{2,1}}}^{{Z_0}} \oplus  \cdots  \oplus {P_{0,{s_{{N_s},1}}}}A_{{s_{{N_s},1}}}^{{Z_0}}} \right){B_1} +  \\
  {\text{ }}\left( {{P_{0,{s_{1,2}}}}A_{{s_{1,2}}}^{{Z_1}} \oplus {P_{0,{s_{2,2}}}}A_{{s_{2,2}}}^{{Z_1}} \oplus  \cdots  \oplus {P_{0,{s_{{N_s},2}}}}A_{{s_{{N_s},2}}}^{{Z_1}}} \right){B_2} +  \cdots  +  \\
  {\text{ }}\left( {{P_{0,{s_{1,{N_p}}}}}A_{{s_{1,{N_p}}}}^{{Z_{{N_p} - 1}}} \oplus {P_{0,{s_{2,{N_p}}}}}A_{{s_{2,{N_p}}}}^{{Z_{{N_p} - 1}}} \oplus  \cdots  \oplus {P_{0,{s_{{N_s},{N_p}}}}}A_{{s_{{N_s},{N_p}}}}^{{Z_{{N_p} - 1}}}} \right){B_{{N_p}}} \\
\end{gathered}
\end{equation}
where $B_j$ is defined in (\ref{eq2}) and $A_{{s_{i,1}}}^{{Z_0}} \triangleq 1{\text{~,}}\forall i
\leqslant {N_p}$. Now, the optimal SM is found by solving the
following optimization problem,
\begin{equation}\label{eq6}
{S^*} = \mathop {\arg \max }\limits_{{s_{1,1}},{s_{1,2}}, \ldots ,{s_{{N_s},{\text{~}}{N_p}}}} {\text{~Q}}
\end{equation}

Based on the derived optimization problem, we can find the optimal
SM by exploiting exhaustive search. Assume that the computational complexity
of computing (\ref{eq5}) for a given SM, $S$, is in
$\mathcal{O} \left( 1 \right) $.
Then, the computational complexity of finding the optimal SM is in
$\mathcal{O} \left( {{N_p}^{{N_p} \times {N_s}}} \right)$.
Since the expression describing the performance metric (the CRN
throughput) is complicated in general, there is no much room for
solving (\ref{eq6}) through classical optimization procedures. On
the other hand, solving (\ref{eq6}) through the
exhaustive search makes no guarantee for fairness among the
SUs. In addition, it results in massive computational burden which is
not scalable regarding to both $N_p$ and $N_s$. All these facts
make a strong motivation and interest in developing an appropriate
suboptimal solution for the problem formulated in (\ref{eq6}).
In the following sections, we propose suboptimum solutions for the SM for three different cases. Advantages of proposed algorithms are threefold: First, it offers
low computational complexity. Second, it provides fairness among
the SUs. Finally, its consumed sensing energy to find a
transmission opportunity is much less than the
exhaustive search.

\section{SMS Algorithm}\label{section4}

\subsection{Structure of the SMS Algorithm}

The proposed algorithm, designed for error-free sensing case, is composed of $N_p$ rounds. In the
$k$-th round, the coordinator determines the $k$-th column of the
SM, i.e., $s_{i,k},~1 \le i \le {N_s}$.
As mentioned before, repeating
a spectrum in the SM for more than one times either in the same mini-slot or in different mini-slots
does not have any benefits on the network throughput.
During each round and for each SU, the coordinator assigns a reward to each
candidate channel\footnote{the channel that has not been assigned to the SS of any user previously} to be possibly allocated to the SS of the SU at that round and then adopts the channel with the maximum reward.
That is, at the round $m$, for each secondary user $k$ and for each unassigned
channel $i$, we define $G_i^{\left( k \right)}\left( m \right)$ as the reward of the channel $i$ if selected as the $m$-th component of the $k$-th SU's sensing sequence.
This reward is set equal to the contribution of the $k$-th
SU to the network throughput if the $i$-th channel is selected, as will be described latter.
Then, the channel with the maximum reward is selected.

We denote the set of all assigned channels to the sensing matrix by $A$. At the
beginning, we have $A = \emptyset $ and $S = \emptyset $ ,where $S$ is the sensing matrix and $\emptyset $ denotes empty matrix. We also denote the set of all channels by $\mathbb{N}$, where $\mathbb{N} = \left\{ {1,2,
\ldots ,{N_p}} \right\}$.
The process is as follows:

\textbf{Round-1}

For this round, first the coordinator assigns a spectrum to the SS of the first SU at its first mini-slot. The coordinator must adopt $s_{1,1}$ from the unassigned channels, i.e., $\overline A  = \mathbb{N}\backslash A = \mathbb{N} $.
We have $G_i^{\left( 1 \right)}\left( 1 \right) = {P_{0,i}}{B_1}$, where $B_1$ is defined in (\ref{eq2}). The coordinator selects a channel with the highest reward for the $s_{1,1}$. That is, the first channel to be sensed by the first SU is,
\begin{equation}\label{eq7}
{s_{1,1}} = \mathop {\arg \max }\limits_{i \in \overline A }
{\text{~}}G_i^{\left( 1 \right)}\left( 1 \right)
\end{equation}

After $ s_{1,1} $ is determined, $A$ and $\overline A $ are respectively updated to $A = \left\{
{{s_{1,1}}} \right\}$ and $\overline
A  = \mathbb{N}\backslash A$. This procedure is repeated for each SU; so for the $\ell$-th user in the first round, we have,
\begin{equation}\label{eq8}
{s_{\ell,1}} = \mathop {\arg \max }\limits_{i \in \overline A }
{\text{~}}G_i^{\left( \ell \right)}\left( 1 \right)
\end{equation}
where $ G_i^{\left( \ell \right)}\left( 1 \right) = {P_{0,i}}{B_1} $.

\textbf{Round-m}

At the $m$-th round, for each SU, the coordinator similarly assigns a reward to
each left spectrums and allocates the best spectrum, which
has the maximum reward, to the SU.
If the coordinator chooses the $j$-th channel for the $m$-th sensing mini-slot of the $\ell$-th SU, the following reward will be gained by the user.
\begin{equation}\label{eq10}
G_j^{\left( \ell  \right)}\left( m \right) = \left(
{\prod\limits_{i = 1}^{m - 1} {\left( {1 - {P_{0,{s_{\ell ,i}}}}}
\right)} } \right){P_{0,j}}{B_m}
\end{equation}

Hence, the coordinator determines the $m$-th element of the SS of
the $\ell$-th SU as,
\begin{equation}\label{eq11}
{s_{\ell ,m}} = \mathop {\arg \max }\limits_{i \in \overline A }
{\text{~}}G_i^{\left( \ell  \right)}\left( m \right)
\end{equation}

At this round, first
it must be determined from which SU the procedure should be started. In order to achieve an acceptable level of fairness among the SUs, the algorithm starts with a SU that has gained the lowest cumulative rewards during the previous $ \left( m - 1 \right) $ rounds
(previous $ \left( m - 1 \right)$ mini-slots). The cumulative
reward gained during the previous $ \left( m - 1 \right)$
mini-slots is calculated for the $\ell$-th SU as,
\begin{equation}\label{eq9}
 \sum\limits_{k = 1}^{m - 1} {G_{{s_{\ell,k}}}^{\left( \ell \right)}\left( k \right)}  = G_{{s_{\ell,1}}}^{\left( \ell \right)}\left( 1 \right) + G_{{s_{\ell,2}}}^{\left( \ell \right)}\left( 2 \right) +  \cdots G_{{s_{\ell,m - 1}}}^{\left( \ell \right)}\left( {m - 1} \right)
\end{equation}
where ${G_{{s_{\ell,k}}}^{\left( \ell \right)}\left( k \right)}$ is defined in (\ref{eq10}).


This process continues until $\left| A \right| = {N_p}$ or
equivalently $\overline A  = \emptyset $. At the end of the
process, the elements of $S$ without any assigned spectrum are replaced
by zero, which indicates that the sensing is not performed for those
elements. Since each channel is sensed only once in the proposed algorithm, the energy consumed by the
SMS algorithm equals to $N_p E_c$ at the worst case, and it does
not increase by $N_s$.

In order to have mid-term fairness, at the beginning of second
time-slot, the process starts with the second SU and the first element of the sensing sequence of this user is determined and then the procedure is continued by selecting the first element of the third user, and at the last the first element of the first user is selected. The other elements are determined as described above. This cyclic ordering is
continued in the following time-slots, i.e., at the beginning of $m$-th run
of the SMS procedure ($m$-th time-slot) the process starts with selecting
the first element of the sensing sequence of the $k$-th SU, where
\begin{equation}\label{eq26}
k = \bmod \left( {m,{N_p}} \right)
\end{equation}

These procedures are summarized in Algorithm \ref{alg1}.
\begin{algorithm}\label{alg1}
\caption{sensing matrix setting algorithm for perfect sensing case(the SMS algorithm)}
\label{alg1}
\begin{algorithmic}
\STATE \textbf{Initialization: }$S = \emptyset $, $\mathbb{N} = \left\{ {1,2, \ldots ,{N_p}} \right\}$, $ A = \emptyset$, $\overline A  = \mathbb{N}\backslash A $, and $m = 1$
\STATE $H \leftarrow $ sort the SUs based on their numbers, sequentially.
\FOR{$m=1$ to (the maximum number of the SUs' mini-slots)}
\FOR{$\ell=$ (the first element of $H$) to (the last element of $H$)}
\FOR{$i=1$ to $N_p$}
\IF{$\overline A \ne \emptyset $}
\STATE Compute $G_i^{\left( \ell  \right)}\left( m \right)$ as in (\ref{eq10})
\STATE Assign ${s_{\ell ,m}} \leftarrow \mathop {\arg \max }\limits_{i \in \overline A } {\text{~}}G_i^{\left( \ell  \right)}\left( m \right)$

\ELSE
\STATE ${s_{\ell ,m}} \leftarrow 0 $

\ENDIF

\ENDFOR
\IF{${\text{the~spectrum~}}i{\text{~is~assigned~to~}}{s_{\ell ,m}}$}
\STATE Add $\left\{ i \right\}$ to $A$
\STATE Update $\overline A$
\ENDIF

\ENDFOR

\STATE $H \leftarrow $ sort the SUs based on their
cumulative rewards computed using (\ref{eq9}).

\ENDFOR
\STATE Return $S$ as the suboptimal sensing matrix.
\end{algorithmic}
\end{algorithm}


\subsection{Computational Complexity}
As we stated before, the computational complexity of finding
the optimal SM is in order of $\mathcal{O} \left( {{N_p}^{{N_p}
\times {N_s}}} \right)$, while it is in order of $\mathcal{O}
\left( 1 \right)$ for our proposed method. In the SMS algorithm, a
channel will be assigned to the SM if it offers the
highest reward, defined in (\ref{eq10}), among the left channels. From (\ref{eq10}), it can be easily shown that ${G_j}^{(l)}\left( m \right) > {G_k}^{(l)}\left( m \right)$ if ${P_{0,j}} > {P_{0,k}}$
\footnote{The reason for defining the award as in (\ref{eq10}) is that it can be easily modified
to the non-error-free sensing case and also for the case of
considering different MAC schemes.}. So for the the error-free case,
the information required to determine the SM is the primary-free probabilities of
the channels.

\subsection{Averaged Consumed Energy for Finding a Transmission Opportunity}
Let $ E_c \left( \tau \right) $ and $ E_c \left( \tau_{ho} \right)
$ denote the consumed energy for sensing of each primary channel
and the consumed energy for each HO, respectively. Hence, the average consumed energy
for finding a transmission opportunity can be calculated as,
\begin{equation}\label{eq12}
\left( {N_s + {{\bar g}_1} + {{\bar g}_2} +  \cdots  + {{\bar
g}_{{N_s}}}} \right){E_c}\left( \tau  \right) + \left( {{{\bar
g}_1} + {{\bar g}_2} +  \cdots  + {{\bar g}_{{N_s}}}}
\right){E_c}\left( {{\tau _{ho}}} \right)
\end{equation}
where ${{{\bar g}_i}}$ denotes the average number of HOs required by the
$i$-th SU to find an idle channel.

The processes of channel sensing and signal transmission consume more energy compared to the HO. Therefore, it is rational to ignore the second term $ \left(
{{{\bar g}_1} + {{\bar g}_2} +  \cdots  + {{\bar g}_{{N_s}}}}
\right){E_c}\left( {{\tau _{ho}}} \right) $ in (\ref{eq12}) compared to the first one.

To evaluate the average number of HOs of the $i$-th SU, ${{{\bar g}_i}}$, we consider two following cases: 1) The SU searches among the channels, finds a transmission opportunity, and then transmits, 2) The SU searches among the available channels, but does not find any free channel. Then, ${{{\bar g}_i}}$ can be easily calculated as,
\begin{equation}\label{neweq12}
{\overline g _i } = {\underbrace{{P_{1,s_{i,1}}}{P_{0,s_{i,2}}} + 2{P_{1,s_{i,1}}}{P_{1,s_{i,2}}}{P_{0,s_{i,3}}} +  \cdots  + \left( N_p - 1 \right) P_{0,s_{i,N_p}}  \prod\limits_{j = 1}^{N_p -1}  {P_{1,s_{i,j}}}}_{T1}} + N_p \prod\limits_{j = 1}^{N_p}  {{P_{1,s_{i,j}}}}
\end{equation}
where the term $T1$ represents the average sensed channels by the $i$-th SU until the user finds a transmission opportunity, and the last term demonstrates the case that the SU senses all channels busy, and therefore it does not transmit on any channels assigned to its SS. By substituting (\ref{neweq12}) in (\ref{eq12}), the total average consumed energy for the exhaustive search method is derived as follows:
\begin{equation}\label{eq13}
\sum\limits_{i = 1}^{{N_S}} {{{\bar g}_i}} {E_C} = {E_C}\sum\limits_{i = 1}^{{N_S}} {\left( 1 + \sum\limits_{k = 2}^{{N_P}} {\left( {\left( {k - 1} \right)\prod\limits_{j = 1}^{k - 1} {{P_{1,{s_{i,j}}}}} } \right){P_{0,{s_{i,k}}}}}  \right)}  + {E_C}\sum\limits_{i = 1}^{{N_S}} { {N_p}\prod\limits_{j = 1}^{{N_P}} {{P_{1,{s_{i,j}}}}} }
\end{equation}

On the other hand, for the SMS algorithm, analytically deriving the average consumed energy is complicated. Hence, we only focus on two special extreme case, i.e., the maximum and minimum consumed energies.
For the worst case, which consumes the maximum energy, all the channels appeared in the SM are sensed.
In this case, the consumed energy equals
to ${N_p}{E_C}\left( \tau  \right)$.
For the best case, which consumes the minimum energy, each SU finds the first channel of its SS free and does not need to sense the rest. In this case, the consumed energy is equal to $\min \left( {{N_p},{N_s}}
\right){E_C}\left( \tau  \right)$. It is worth noting that for
${P_{0,j}} = 1 ~,for~all~j$, the sequential
sensing scheme forces the SUs to continue searching among
all channels in their sensing sequences, which
is equivalent to the worst case, and similarly ${P_{0,j}} = 0 ~,for~all~j$ is equivalent to the best case with minimum consumed energy. If we compute the average consumed energy of the optimum solution given in (\ref{eq13}) for theses two  cases, the maximum and minimum consumed energies will be equal to ${E_C}{N_s}\left( {1 + {N_p}} \right)$ and ${E_C}{N_s}$, respectively, which are higher than those of the SMS algorithm.

For more elaboration, we study an especial case where all channels have
the same primary-free probabilities, i.e., ${P_{0,j}} =
P{\text{~}},{\text{~~}}{P_{1,j}} = 1 - P{\text{~}}
,for~all~j $. Then, we can simplify (\ref{eq13}) as follows,
\begin{equation}\label{eq14}
\begin{gathered}
  {E_C}\left( \tau  \right)\sum\limits_{i = 1}^{{N_S}} {{{\bar g}_i}}  = {E_C}\left( \tau  \right)\left( {{N_S}P + P\sum\limits_{i = 1}^{{N_S}} {\left( {\sum\limits_{k = 2}^{{N_p}} {\left( {\left( {k - 1} \right){{\left( {1 - P} \right)}^{k - 1}}} \right)} } \right) + } {N_S}{N_p}{{\left( {1 - P} \right)}^{{N_p}}}} \right) \\
   = {E_C}\left( \tau  \right){N_S}\left( {P + \frac{{1 - P + \left( { - 1 + P - P{N_p}} \right){{\left( {1 - P} \right)}^{{N_p}}}}}{P} + {N_p}{{\left( {1 - P} \right)}^{{N_p}}}} \right) \\
   = {E_C}\left( \tau  \right){N_S}\left( {\frac{{{{\left( {1 - P} \right)}^2} - {{\left( {1 - P} \right)}^{{N_p} + 1}}}}{P} + 1} \right) \\
\end{gathered}
\end{equation}

It is worth noting that (\ref{eq14}) is a decreasing function of $P$. Hence, the minimum and maximum values of consumed energy are related to the cases $P = 0$ and $P = 1$, as discussed above. Moreover, as will be shown in the numerical result section (Fig.~\ref{fig6}), the consumed energy associated with the optimal SM is higher than the consumed energy for the matrix obtained by the SMS algorithm for all values of $P$.

In the following, the impacts of the sensing errors are investigated.
In general, the sensing error manifests itself in two forms: false
alarm and miss-detection.
In the SMS algorithm proposed, a channel is
allocated to only one SS, and thus the SUs have no common channels
in their sequences. Although this approach performs well when
there is no sensing error, but in the case of non-perfect sensing,
this method is not efficient; since by a false alarm made by a SU in a sensed
channel, a transmission opportunity is
lost for this channel by all SUs. Therefore, it
seems that the coordinator has to repeat spectrums in the $S$ in
order to increase the possibility of exploiting all opportunities and thus to increase the spectral efficiency.
On the other hand, allocating a channel to the sensing sequences of
multiple SUs increases the average number of sensed channels and
thus raises the average sensing energy consumption. Moreover, due to miss-detection, it is possible for a SU to mistakenly transmit on a channel which is already used by another SU or PU, and therefore some collisions
might occur. Hence, there is a trade-off between the
average achievable throughput, energy consumption, and the level
of collision in the CRN which must be addressed in an extension
of the SMS algorithm.

To modify the SMS algorithm, we must consider the impact of sensing
error probabilities on the reward function. Moreover,
the channels occupation probabilities will be different at the beginning of the
various mini-slots, which must be reflected in the
reward function. Finally, because of repetition of each channel in
the SM, the stopping rule, which was $\left| A \right| = {N_p}$,
must be modified.

\section{MSMS Algorithm}\label{section5}

Since in the SMS algorithm, the sensing sequences of the SUs have no common channels, the occupancies of channels at the beginning of each mini-slot only depend on the PUs' activities. But if the channels are allowed to be repeated
in multiple rows or columns of the SM, the occupancy of a channel can
be due to the presence of either the PU or a SU. To extend the SMS algorithm, first the occupation probability of the $j$-th channel, i.e., ${q_{1,j}}$, has to be determined.

It is worth noting that, since all the SUs use the same sensing
schemes with the same sensing time lengths, they
all have the same probabilities of false alarm and miss-detection.
Thus, we have,
\begin{equation}\label{eq15}
\begin{gathered}
  {P_{fa,1}} = {P_{fa,2}} =  \cdots  = {P_{fa,{N_p}}} = {P_{fa}} \hfill \\
  {P_{d,1}} = {P_{d,2}} =  \cdots  = {P_{d,{N_p}}} = {P_d} \hfill \\
\end{gathered}
\end{equation}

\begin{figure}[h]
  \centering
  \includegraphics[width= 3.5 in]{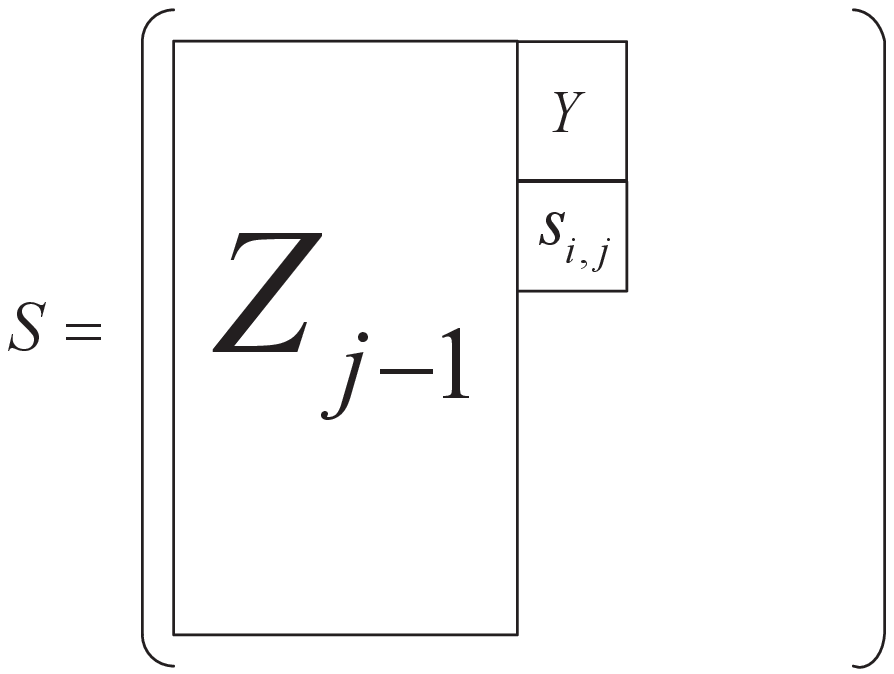}\\
  \caption{Graphical representation of the sensing matrix.}
  \label{fig2}
\end{figure}

In order to reflect the impact of sensing error on the proposed
algorithm, three possible cases must be considered when the
coordinator tends to adopt the $l$-th channel as $s_{i,j}$:
\begin{itemize}
  \item  The $l$-th channel has not yet been allocated to any sensing sequences, i.e., $l \notin {Z_{j - 1}}$ (matrix Z is defined in section \ref{section3} and Fig.~\ref{fig2}).
  \item  The $l$-th channel has been adopted at least once for sensing at the previous mini-slots, i.e., $l \in {Z_{j - 1}}$.
  \item  The $l$-th channel has been allocated simultaneously to multiple users at the $j$-th mini-slot (vector $Y$ shown in Fig. 2).
\end{itemize}

Fig.~\ref{fig2} depicts these cases graphically. Suppose that
$j$-th component of the SS of the $i$-th SU, i.e., $s_{i,j}$, is to be selected by the
coordinator. So the reward gained by adopting the $k$-th channel
as $s_{i,j}$ is to be determined. Considering the definition of the matrix $Z$ and the vector $Y$ in Section \ref{section3} and also in Fig.~\ref{fig2}, if $n$ elements of $Z$ are equal to
$k$, this will indicate that the $k$-th channel has been sensed at most by $n$ SUs
during previous mini-slots. Also, if two or more elements of $Y$
are equal to $k$, then the $k$-th channel will be sensed by two or
more SUs during the $j$-th mini-slot. When channels are allowed to
be sensed by multiple SUs simultaneously, an appropriate MAC
protocol can be used to regulate the access of the SUs to
transmission opportunities. As a first step, we assume that a SU
starts transmitting when it finds a transmission opportunity.
Applying more appropriate MAC protocols to decrease the collision probability among the SUs will be considered
later. For the mentioned transmission policy, if the $k$-th channel belongs to $Y$ and it is
also adopted as $s_{i,j}$ by the coordinator, a collision may
occur and the reward may be zero. Without loss of generality, it
is assumed that the coordinator starts the allocation process for
the $j$-th mini-slot from the top of the $j$-th column of matrix
$S$.

Given $Z_{j-1}$, we denote the occupation probability of the $k$-th channel at the beginning of the $j$-th mini-slot as
$q_{1,k}^{\left( j \right)}$. Then, we easily obtain,
\begin{equation}\label{eq16}
q_{1,k}^{\left( j \right)} = {P_{1,k}} + {\theta _{1,k,{Z_{j - 1}}}}
\end{equation}
where $\theta _{1,k,{Z_{j - 1}}}$ represents the probability of several sensing of and possibly transmitting on the $k$-th channel in the first $ \left( j - 1 \right)$ mini-slots and is easily computed as,
\begin{equation}\label{eq17}
{\theta _{1,k,{Z_{j - 1}}}} = \sum\limits_{n = 1}^{\left| {{Z_{j - 1}}} \right|} {\delta _n^{k,{Z_{j - 1}}}\left( {1 - {P_{0,k}}{P_{fa}}^n} \right)}
\end{equation}
where $\left| {{Z_{j - 1}}} \right| = {N_s}\left( {j - 1} \right)$ is the number of elements of $Z_{j - 1}$ and $\delta _n^{k,{Z_{j - 1}}}$ is defined as,
\begin{equation}\label{eq18}
\delta _n^{k,{Z_{j - 1}}} = \left\{ {\begin{array}{*{20}{c}}
   {1{\text{~:~~if~spectrum~}}\# k{\text{~is~in~}}Z_{j-1}{\text{~for~}}n{\text{~times}}} \hfill  \\
   {0{\text{~:~~if~spectrum~}}\# k{\text{~is~not~in~}}Z_{j-1}{\text{~for~}}n{\text{~times}}} \hfill  \\
\end{array} } \right.
\end{equation}


As in the SMS algorithm, the process starts with $S = \emptyset $
and at the first step $s_{1,1}$ is selected for the SS of the first SU by the
coordinator. As before, for channel $i \in \mathbb{N}$, $G_i^{\left( k
\right)}\left( 1 \right)$ denotes the reward contributed by the
$k$-th SU to the overall throughput of the secondary network when
the $i$-th channel is allocated as the first element of its
sensing sequence.

\textbf{Round-1}

At the first round of the MSMS algorithm, $G_i^{\left( 1
\right)}\left( 1 \right) = q_{0,i}^{\left( 1 \right)}\left( {1 -
{P_{fa}}} \right){B_1}$ in which $B_1$ is defined in (\ref{eq2}).
$q_{0,i}^{\left( 1 \right)} = 1 - q_{1,i}^{\left( 1 \right)}$, where $q_{1,i}^{\left( 1 \right)}$ is given in (\ref{eq16}). Therefore, $s_{1,1}$ is determined as,
\begin{equation}\label{eq19}
{s_{1,1}} = \mathop {\arg \max }\limits_{i \in \mathbb{N}} {\text{ }}G_i^{\left( 1 \right)}\left( 1 \right)
\end{equation}

If the $i$-th spectrum is adopted as $s_{l,1}$, the reward added
to the system can be calculated as
\begin{equation}\label{eq20}
G_i^{\left( l \right)}\left( 1 \right) = \sum\limits_{n = 0}^{\left| Y \right|} {\delta _n^{i,Y}q_{0,i}^{\left( 1 \right)}{P_{fa}}^n\left( {1 - {P_{fa}}} \right){B_1}}
\end{equation}
in which
\begin{equation}\label{eq21}
\delta _n^{i,Y} = \left\{ {\begin{array}{*{20}{c}}
   {1{\text{~:~~if spectrum~}}\# i{\text{~is~in~}}Y{\text{~for~}}n{\text{~times}}} \hfill  \\
   {0{\text{~:~~if spectrum~}}\# i{\text{~is~not~in~}}Y{\text{~for~}}n{\text{~times}}} \hfill  \\
\end{array} } \right.
\end{equation}

Finally, the first channel of the SS of the $l$-th SU is selected according to:
\begin{equation}\label{eq22}
{s_{\ell ,1}} = \mathop {\arg \max }\limits_{i \in \mathbb{N}} {\text{ }}G_i^{\left( \ell  \right)}\left( 1 \right)
\end{equation}

\textbf{Round-m}

The $l$-th SU gains a reward by adopting the $j$-th channel as the
$m$-th element of its sensing sequence provided that the user has not
detected a transmission opportunity in its previous sensed channels. Note that besides finding a truly free channel, the SU may mistakenly sense an occupied channel as free due to miss detection and does not
continue the sensing procedure. Therefore, the reward gained by the $l$-th SU is:
\begin{equation}\label{eq23}
G_j^{\left( \ell  \right)}\left( m \right) = \left( {\underbrace {\prod\limits_{i = 1}^{m - 1} {\left( {q_{0,{s_{\ell ,i}}}^{\left( i \right)}{P_{fa}} + q_{1,{s_{\ell ,i}}}^{\left( i \right)}{P_d}} \right)} }_{C1}} \right)\underbrace {\sum\limits_{n = 0}^{\left| Y \right|} {\delta _n^{j,Y}q_{0,j}^{\left( m \right)}{P_{fa}}^n\left( {1 - {P_{fa}}} \right){B_m}} }_{C2}
\end{equation}
where $C1$ indicates the probability of requiring $\left( m - 1 \right)$ HO,
and $C2$ represents the throughput contribution of $j$-th channel if selected at the
$m$-th mini-slots of the $l$-th SU for the transmission.

Thus the $m$-th element of the $l$-th sensing sequence is similarly
determined as,
\begin{equation}\label{eq24}
{s_{\ell ,m}} = \mathop {\arg \max }\limits_{j \in \mathbb{N}} {\text{ }}G_j^{\left( \ell  \right)}\left( m \right)
\end{equation}

Similar to the SMS scheme, in the MSMS algorithm, at the round-m $\left( {m \in \left\{ {2,3,
\ldots ,{N_p}} \right\}} \right)$ the coordinator starts with the
SU that has gained less cumulative rewards in its $\left( m - 1
\right)$ previous mini-slots. The
cumulative rewards of the $j$-th SU at its $\left( m - 1 \right)$
previous mini-slots can be computed as,
\begin{equation}\label{eq25}
G_{{s_{j,1}}}^{\left( j \right)}\left( 1 \right) +
G_{{s_{j,2}}}^{\left( j \right)}\left( 2 \right) +  \cdots
G_{{s_{j,m - 1}}}^{\left( j \right)}\left( {m - 1} \right)
\end{equation}
where $G_{{s_{j,i}}}^{\left( j \right)}\left( i \right)$ for $1
\leqslant i \leqslant m - 1$ is calculated as (\ref{eq23}). Hereby,
a certain level of fairness is ensured among the SUs.

The stopping rule of the MSMS algorithm is different from that of
the SMS algorithm. For the MSMS algorithm, two possible rules can
be exploited. First, there exist no constraint on the number of
times that each channel can be used as the elements of the SM. For this case, the
process is stopped when all the elements of the SM have been
selected. Second, the number of times that each channel is
appeared in the SM is limited. While the first rule leads to the
maximum average throughput which can be achieved by the MSMS
algorithm, the second rule is more rational and practical. The
probability of a channel erroneously sensed as busy exponentially
decreases by the number of times that the channel is sensed.
As a result, we use the second stopping
rule. In the numerical result part, we limit the number of times that each channel is appeared
in the SM to $3$. That is, the coordinator assigns each channel, if necessary, at most
three times in the SM.

In order to have further mid-term fairness among the SUs, the same
idea as applied to the SMS algorithm is exploited, i.e., at the beginning of
$m$-th run of the MSMS procedure ($m$-th time slot), the process starts with the $k$-th SU as specified in
(\ref{eq26}). The procedures of MSMS algorithm are summarized
in Algorithm \ref{alg2}.
\begin{algorithm}\label{alg2}
\caption{sensing matrix setting algorithm for non-perfect sensing
case(the MSMS algorithm)} \label{alg2}
\begin{algorithmic}
\STATE \textbf{Initialization: }$S = \emptyset $, $\mathbb{N} =
\left\{ {1,2, \ldots ,{N_p}} \right\}$, $ \overline A =
\mathbb{N}$, $m = 1$, and $RN = {\left\{ {0,0, \ldots ,0}
\right\}_{1 \times {N_p}}}$ \STATE $H \leftarrow $ sort the SUs
based on their numbers, sequentially. \FOR{$m=1$ to (the maximum
number of the SUs' mini-slots)} \FOR{$\ell=$ (the first element of
$H$) to (the last element of $H$)} \FOR{$i=1$ to $N_p$}
\IF{$\overline A \ne \emptyset $} \STATE Compute $q_{1,i}^{\left(
m \right)}$ using (\ref{eq16}), (\ref{eq17}), and (\ref{eq18})
\STATE Compute $G_i^{\left( \ell  \right)}\left( m \right)$ as in
(\ref{eq23}) \STATE Assign ${s_{\ell ,m}} \leftarrow \mathop {\arg
\max }\limits_{i \in \overline A } {\text{~}}G_i^{\left( \ell
\right)}\left( m \right)$

\ELSE
\STATE ${s_{\ell ,m}} \leftarrow 0 $

\ENDIF

\ENDFOR
\IF{${\text{the~spectrum~}}i{\text{~is~assigned~to~}}{s_{\ell ,m}}$}
\STATE $RN\left[ i \right] \leftarrow \left( RN\left[ i \right] + 1 \right)$
\ENDIF

\IF{$RN\left[ i \right] = 3$}
\STATE $\overline A = \mathbb{N} \backslash \left\{ i \right\}$
\ENDIF

\ENDFOR

\STATE $H \leftarrow $ sort the SUs based on their
cumulative rewards computed using (\ref{eq25}).

\ENDFOR
\STATE Return $S$ as the suboptimal sensing matrix.
\end{algorithmic}
\end{algorithm}



\section{PMSMS Algorithm}\label{section6}

Regardless of how the SM is created, it is possible for a channel to
be assigned to the several SUs in the same mini-slot. In this
case, various conventional MAC algorithms can be exploited to
increase  the transmission chance on this channel. In this section, we
utilize the well-known p-persistent MAC (PMAC) protocol in the MSMS algorithm
and develop PMSMS algorithm. In this algorithm, in each mini-slot
the SUs sense the assigned channels with the probability of $p$.
In order to have a synchronous sensing scheme for all SUs, the SU
will be idle for $\tau$ seconds(mini-slot time duration) if its MAC protocol does not allow
it to sense the channel.

The stopping rule as well as fairness establishment techniques are
similar to the MSMS algorithm. Considering PMAC, there are two
cases that a free channel is not used by a SU. First case is due
to the false alarm, and the second is due to the presence of PMAC
protocol. In the latter case, the channel is not sensed with the
probability of $ \left( 1 - p \right) $. Considering these two
cases easily leads to the following modification of ${\theta_{1,k,{Z_{j - 1}}}}$
(defined in (\ref{eq17})).
\begin{equation}\label{eq27}
{\theta _{1,k,{Z_{j - 1}}}} = \left\{ {\begin{array}{*{20}{c}}
   {\sum\limits_{n = 1}^{\left| {{Z_{j - 1}}} \right|} {\delta _n^{k,{Z_{j - 1}}}\left( {1 - \sum\limits_{t = 0}^n {\left( {\begin{array}{*{20}{c}}
   n  \\
   t  \\
\end{array} } \right){P_{0,k}}{P_{fa}}^t{{\left( {1 - {P_{fa}}} \right)}^{n - t}}{{\left( {1 - p} \right)}^{n - t}}} } \right),} } \hfill & {0 \leq p < 1} \hfill  \\
   {\sum\limits_{n = 1}^{\left| {{Z_{j - 1}}} \right|} {\delta _n^{k,{Z_{j - 1}}}\left( {1 - {P_{0,k}}{P_{fa}}^n} \right)} ,} \hfill & {p = 1} \hfill  \\
\end{array} } \right.
\end{equation}

Then, the channel occupation probability is obtained by substituting
(\ref{eq27}) in (\ref{eq16}). The generalized reward of assigning
the $j$-th primary channel to the SS of the $\ell$-th SU at the
$m$-th mini-slot is simply calculated as,
\begin{equation}\label{eq28}
G_j^{\left( \ell  \right)}\left( m \right) = \left( {\prod\limits_{i = 1}^{m - 1} {\left( {q_{0,{s_{\ell ,i}}}^{\left( i \right)}{P_{fa}} + q_{1,{s_{\ell ,i}}}^{\left( i \right)}{P_d}} \right)} } \right) \times \sum\limits_{n = 0}^{\left| Y \right|} {\delta _n^{j,Y}q_{0,j}^{\left( m \right)}{B_m}} \hbar
\end{equation}
where
\begin{equation}\label{hbareq28}
\hbar  = \left\{ {\begin{array}{*{20}{c}}
   {\left( {1 - {P_{fa}}} \right)p\sum\limits_{k = 0}^n {\left( {\begin{array}{*{20}{c}}
   n  \\
   k  \\
\end{array} } \right){P_{fa}}^k{{\left( {1 - {P_{fa}}} \right)}^{n - k}}{{\left( {1 - p} \right)}^{n - k}}} ,} \hfill & {0 \leq p < 1} \hfill  \\
   {\left( {1 - {P_{fa}}} \right){P_{fa}}^n{\text{ }},} \hfill & {p = 1} \hfill  \\
\end{array} } \right.
\end{equation}
and finally the coordinator adopts a channel with a highest reward for the $m$-th
mini-slot of $l$-th SU as follows,
\begin{equation}\label{eq29}
{s_{\ell ,m}} = \mathop {\arg \max }\limits_{i \in \mathbb{N} } {\text{ }}G_i^{\left( \ell  \right)}\left( m \right)
\end{equation}

The SUs sense the assigned channels with more probability as $p$
increases, which can increase the chance of finding a
transmission opportunity, at the expense of raising the level of contention
among the SUs. Therefore, there is a tradeoff on the value of $p$, which will be discussed in the next section.

\section{Numerical Results}\label{section7}

In this section, the performance of the proposed allocation
schemes is evaluated by simulation considering the effect of
different parameters. Moreover, advantages of exploiting the
proposed algorithms are demonstrated through exhaustive
simulations.

The simulation parameters are given in Table~\ref{table1}. The
values of SNR and sampling frequency which are used by the energy
detector are adopted from \cite{liang2008}. The value of sensing
time of each channel, $ \tau $, is selected such that the false
alarm and detection probabilities meet the constraints imposed by
the \emph{IEEE 802.22} standard \cite{stevenson}. Each SU senses
the channels according to its sensing sequence, each for $\tau$
seconds, until a free channel is found. Then, the SU
transmits on this channel for the rest of the time slot. The
average normalized CRN throughput has been evaluated by simulating
the scenario for $100$ time slots.

\begin{figure}[h]
\hspace{-0.5cm}
  \centering
  \includegraphics[width= 6.5 in]{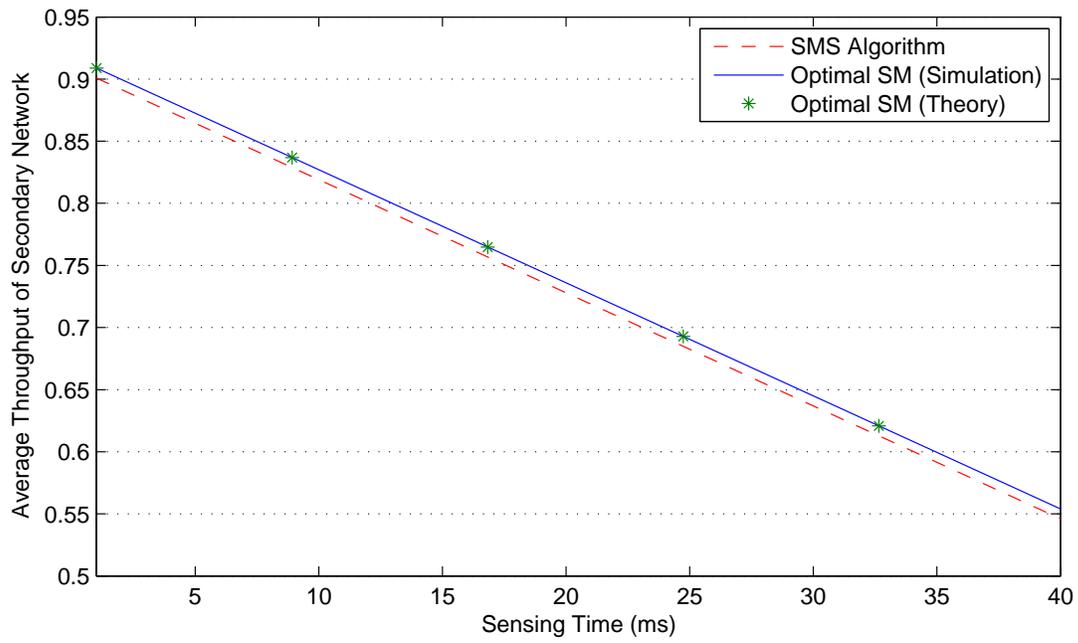}\\
  \caption{Average throughput versus sensing time for various sensing matrix selection schemes}
  \label{fig4}
\end{figure}

Fig.~\ref{fig4} validates our analysis and depicts the average
throughput (normalized to $R$) of the SUs versus the sensing time, for
the optimal SM and the SM obtained based on SMS algorithm, for a
error-free sensing case. For the optimal SM, both the theoretical
and simulation results are provided. As it can be realized, the
throughput linearly decreases by the sensing time. This is due to the
fact that for the error-free case, there exists no error in the
detection scheme and thus while the increase of sensing time does not
have any positive impact on the correct detection, it linearly reduces
the transmission time, as can be inferred from (\ref{eq2}).
Fig.~\ref{fig4} also verifies near-optimality of our proposed
algorithm; while it imposes much less complexity burden than the
optimum scheme. The relative difference between the average
throughput obtained by the SMS algorithm and that obtained by the
exhaustive search method is negligible and about $0.81\%$.

\begin{figure}[h]
\hspace{-0.5cm}
  \centering
  \includegraphics[width= 6.5 in]{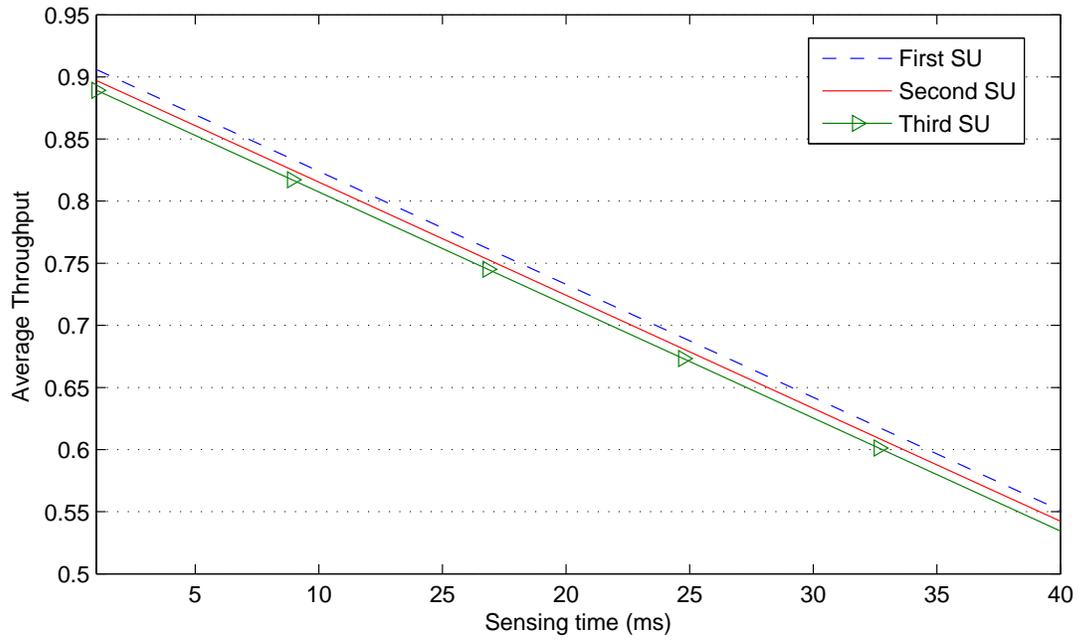}\\
  \caption{Average throughput versus sensing time for different secondary users}
  \label{fig5}
\end{figure}
\begin{figure}[]
\hspace{-0.5cm}
  \centering
  \includegraphics[width= 6.5 in]{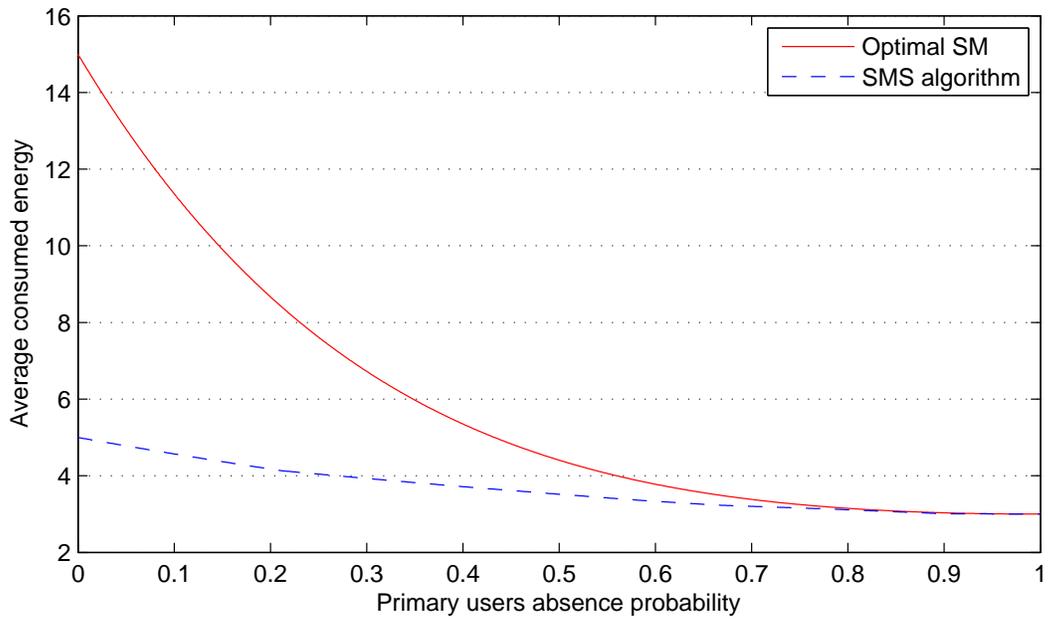}\\
  \caption{Average consumed energy versus the primary user absence probability}
  \label{fig6}
\end{figure}

Fig.~\ref{fig5} compares the average throughputs of different SUs
for a CRN with three SUs, again for a error-free sensing scheme.
In this example, it is assumed that the number of primary
channels, $N_p$, is $5$. The maximum relative difference
between the SUs throughputs is $1.84\%$ which confirms the
fairness among the SUs when using the proposed scheme. It is expected
by running the simulation for more than $100$ times, the
difference among the SUs' throughputs disappears.

Fig.~\ref{fig6} demonstrates the energy efficiency of our proposed
algorithm. This Fig. shows the average consumed energy
versus the primary user occupation
probability. The SUs consumes less average energy to find a
transmission opportunity when sensing the channel based on the SM
obtained by our proposed method compared to the optimal SM. It is
worth noting that in both schemes, the consumed energies of the SUs
increase when the PUs' absence probability decreases; as the SUs have to sense more channels
to find free ones.

\begin{figure}[]
\hspace{-0.5cm}
  \centering
  \includegraphics[width= 6.5 in]{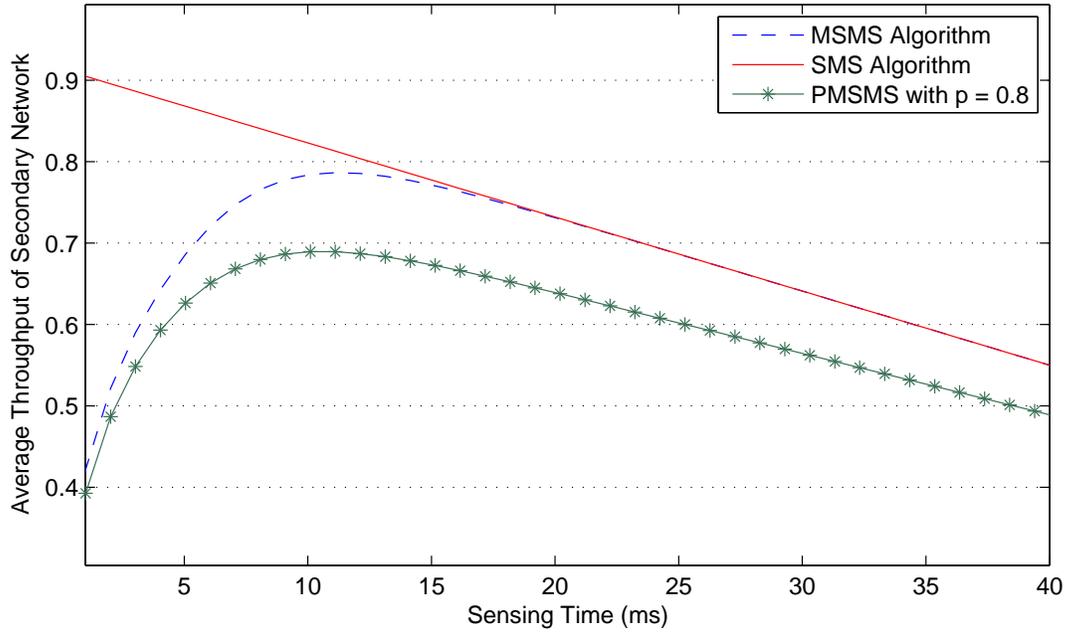}\\
  \caption{Average throughput of SUs versus sensing time for various proposed schemes.}
  \label{fig7}
\end{figure}

Fig.~\ref{fig7} compares the average throughputs of various
spectrum allocation schemes proposed in this paper. For the two
MSMS and PMSMS schemes, a practical scenario with activity detection errors
has been assumed. In general, as the sensing time increases, the
detector senses the channels more accurately and finds more
transmission opportunities. However, by the increase
of the sensing time, less time remains for the transmission.
Hence, there exist a tradeoff between average throughput and
detector accuracy. As it is seen in Fig.~\ref{fig7}, first the
SUs' throughput increases by $\tau$ (due to an accurate sensing);
then after an optimum point, where $P_{md}$ and $P_{fa}$ are in
acceptable levels, the throughput starts decreasing due
to the reduction of the time left for the transmission. For a
sensing time greater than a specific amount (optimum value), the
false alarm and miss-detection probabilities of the detector
becomes negligible, and the allocation procedure of the MSMS
algorithm as well as its performance will be similar to those of
the SMS algorithm, for which a error-free sensing has been
assumed. In the MSMS scheme, the SUs sense channels with the
probability of $\left( 1 - P_{fa} \right)$, and thus some transmission
opportunities are lost as a result of false alarm. This is the reason that the
average throughput of the SUs obtained by the MSMS algorithm is
less than that of the SMS algorithm in which $P_{fa}$ is
assumed to be zero. In the PMSMS algorithm, the applied
p-persistent MAC protocol leads to loss transmission
opportunities, and thus to less average throughput compared to the MSMS when the number of SUs is not too high.

\begin{figure}[]
\hspace{-0.5cm}
  \centering
  \includegraphics[width= 6.5 in]{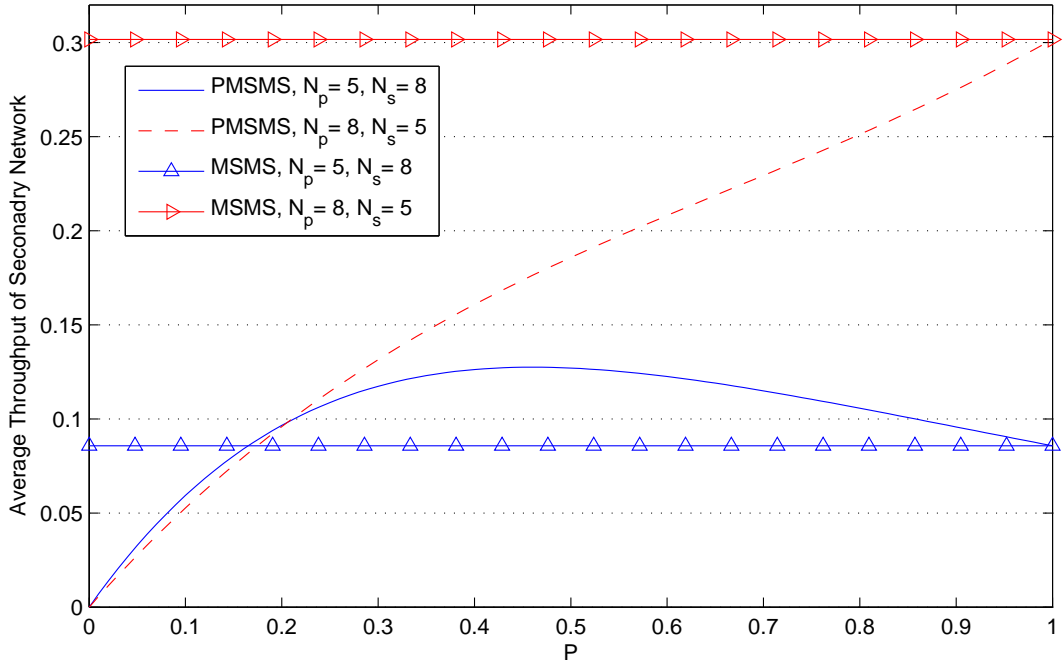}\\
  \caption{Advantages of exploiting PMAC.}
  \label{fig8}
\end{figure}

Fig.~\ref{fig8} demonstrates the advantages of the exploited PMAC for
the case that the number of primary channels are less than the
number of SUs. This Fig. shows the average SUs' throughput
versus the probability of sensing an assigned channel (i.e., $p$
in MAC protocol).
Note that the performance for $p = 1$ is the same as that of the MSMS.
As can be realized, for $N_s=8$ and $N_P = 5$,
the exploited PMAC protocol can increase the chance of transmission
on the channels by reducing the contention level among the SUs. Therefore, the PMSMS scheme can offer higher throughput for the CRN than the MSMS scheme provided that the coordinator appropriately selects the value of $p$, which for the example considered it must be larger than $0.165$. Interestingly, for $ p = 0.46 $, the improvement in the throughput when using the PMSMS scheme is about $ 48.8\% $ compared to the MSMS scheme.

\section{Conclusion}\label{section Conclusion}\label{section8}

In this paper, the average throughput of a cognitive radio network (CRN) for
a given sensing matrix (SM) has been derived, and an optimization
problem has been formulated to find the optimal SM. In order to mitigate the
challenges associated with the optimal solution, three novel centralized suboptimal
algorithms have been proposed. More specifically, the SMS and MSMS schemes are proposed for error-free and non-perfect sensing cases, respectively, and then the PMSMS algorithm is developed by applying the conventional p-persistent MAC protocol in the MSMS scheme to strengthen the multiple access capability of the CRN. Besides offering throughput close to the maximum achievable one, the benefits of these proposed schemes are threefold. In addition to comparatively low computational complexities, they provide an acceptable level of fairness among secondary users. Further, they offer lower consumed energies compared to the optimum solution.
The performance of the proposed
schemes has been evaluated, and their efficiencies have been demonstrated through theoretical analysis as well as exhaustive simulation results.

\bibliographystyle{IEEEtran}
\bibliography{IEEEabrv,Bibliogeraphy_SSS_SMS}
\begin{table}[h]
  \centering
  \caption{Simulation Parameters}\label{table1}
   \begin{tabular}{c c c c c c}
\hline \hline
   Parameter & Description & Value \\ \hline

  $P_d^{\min }$ & Minimum allowable detection probability & 0.9 \\
  $P_{fa}^{\max }$ & Maximum allowable false alarm probability & 0.1 \\
  $f_s$ & Receiver sampling frequency & 6 MHz \\
  $T$ & Time-slot duration & 200 ms \\
  $\tau_{ho}$ & Required time for handover & 0.1 ms \\
  $N_p$ & Number of primary users & 5 \\
  $N_s$ & Number of primary users & 3 \\
  \hline \hline
\end{tabular}
\end{table}

\end{document}